\documentclass[12pt]{article}

\usepackage{amsmath,amsfonts,amssymb,latexsym}

\def\be{\begin{equation}}
\def\ee{\end{equation}}
\def\bq{\begin{eqnarray}}
\def\eq{\end{eqnarray}}
\def\beq{\begin{eqnarray*}}
\def\eeq{\end{eqnarray*}}

\begin{document}
\title{\textsc{Trans-Planckian censorship and spacetime singularities}}
\author{\Large{\textsc{Spiros Cotsakis$^{1,2}$}\thanks{\texttt{skot@aegean.gr}},
 \textsc{John Miritzis}$^3$\thanks{\texttt{imyr@aegean.gr}}}\\
$^1$Institute of Gravitation and Cosmology, RUDN University\\
ul. Miklukho-Maklaya 6, Moscow 117198, Russia\\
$^{2}$Research Laboratory of Geometry,  Dynamical Systems\\  and Cosmology,
University of the Aegean,\\ Karlovassi 83200,  Samos, Greece\\
$^{3}$Department of Marine Sciences\\
University of the Aegean\\
University Hill, Mytilene 81100, Greece}
\date{January 2023}
\maketitle
\begin{abstract}
\noindent We study the effects of  trans-planckian censorship conjecture (TCC) bounds on  geodesic completeness of spacetime and the associated existence for an infinite proper time. Using Gronwall's lemma, TCC bounds can be derived directly, leading to a result about the absence of blowup solutions. We show that the TCC provides part of the required criteria for geodesic completeness, and we then provide the remaining ones - the norm of the extrinsic curvature being bounded away from zero. We also discuss the importance of these results for the classical evolution of Friedmann universes under the assumptions of global and regular hyperbolicity.
\end{abstract}

\section{Introduction}
It is well-known that the Hawking--Penrose theorems provide sufficient conditions for the existence of singularities in spacetime~\cite{he}, while completeness theorems associated with the work of Y. Choquet-Bruhat give sufficient conditions for the possible geodesic completeness of spacetimes~\cite{ycb-c}. In~the first case we have geometric and causality conditions leading to geodesic incompleteness, while in the second case completeness of geodesics is established under various analytic criteria. In~both cases, such conditions may be realized in effective theories and, as~it has been repeatedly emphasized, such theories may not be consistent with modern unification ideas, cf., e.g.,~\cite{tcc0}.

In fact, according to the trans-Planckian censorship conjecture, initial fluctuations can never exit the Hubble radius, and~in this sense  such information can never classicalize and become `visible' to classical evolution~\cite{tcc2,tcc1}. This is like having a cosmological censor that, in~an analogous way to that  in cosmic censorship, hides any trans-Planckian information (see, e.g.,~\cite{tcc3} for more recent work on lower bounds on black hole masses, 
~\cite{tcc4,tcc5} for related work on inflation and dark energy, and~\cite{tcc6} on the influence of negative potentials). Since a central question in studies of the early structure and evolution of the universe is the possible presence of singularities, it is important to understand how the trans-Planckian censorship conjecture relates to the possible  resolution of cosmological~singularities.

The structure of this paper is as follows. In Section 2, 
 we introduce three different forms of trans-Planckian bounds, and~then provide sufficient conditions in the form of integrability assumptions of the Hubble parameter (i.e., extrinsic curvature) that lead to two of them. In~Section~\ref{sec3}, we show how trans-Planckian bounds lead to the absence of a blowup in the classical solutions, and~discuss why such bounds alone cannot provide an overall  criterion for the possible geodesic completeness of spacetime. We then show how one can obtain such criteria by introducing a further condition that we call the `anti-Gronwall assumption', that together with the trans-Planckian bounds may lead to a total bound on the norm of the Hubble parameter. We further discuss these results in Section 5. 

\section{Trans-Planckian~Bounds}
In this Section, we introduce a new method to derive trans-Planckian bounds  based on the Gronwall's~lemma.

We start with  the `Gronwall hypothesis', which is contained in the following differential inequality:
\be \label{hub1}
\frac{\dot{a}(t)}{a(t)}\leq H_0(t),
\ee
for the two functions $a,H_0$ defined for all $t$ in the interval $[t_i,t_f]$ and assumed to be differentiable and nonnegative (weaker assumptions are possible). Using Gronwall's lemma (see~\cite{tao} for a discussion closer to its usage in the present work, and~\cite{mitri,drag} for wider applications and references on inequalities), we find:
\be \label{gr1}
\frac{a(t_f)}{a(t_i)}\leq e^{\int_{t_i}^{t_f}H_0(s)ds}.
\ee

Let 
us first consider the case  that $H_0=\textrm{const}$. For~each finite $t_f$ there is a nonzero constant $H_f$ such that the right-hand side of (\ref{gr1}) is pointwise bounded,  namely,
\be \label{1}
H_0(t_f-t_i)<\ln\frac{M_P}{H_f}.
\ee
Then it follows from the conclusion of the Gronwall's lemma (\ref{gr1}) that:
\be \label{2}
\frac{a(t_f)}{a(t_i)}\,l_P<H_f^{-1},
\ee
with $l_P=M_P^{-1}$ (in other notation, setting $N=H_0(t_f-t_i)$ for the number of `e-folds', if~we assume $e^N<M_P/H_f$ as in (\ref{1}), then (\ref{2}) follows.) We note that the trans-Planckian bound in the form stated in Ref.~\cite{tcc2} does not hold in the interval $[t_i,\infty)$ for each finite $t_i$, because~when the upper endpoint $t_f\rightarrow \infty$, the~left-hand side of (\ref{1}) is~infinite.

We move on to the second case that is when $H_0$ is \emph{not} assumed constant. We suppose that $H_0$ is an integrable function on $[t_i,\infty)$, and~replace the left-hand side of  inequality  (\ref{1}) with the expression $\int_{t_i}^{t_f}H_0(s)ds$. We then  end up with the pointwise assumption that for each $t_f$, we have:
\be \label{1'}
\int_{t_i}^{t_f}H_0(s)ds<\ln\frac{M_P}{H_f}.
\ee
This implies that the statement of the trans-Planckian censorship conjecture as formulated in~\cite{tcc2} now  becomes a trans-Planckian censorship \emph{theorem} provided $H_0$ is integrable:  for 
 any integrable function $H_0(t)$ the integral $\int_{t_i}^{t_f}H_0(s)ds$ is bounded, and~we have:
\be \label{3}
\frac{a(t)}{a(t_i)}\,l_P<H_\infty,\quad t\in[t_i,\infty) ,
\ee
where $H_\infty$ is a suitable constant that provides  a uniform bound for the left-hand side of (\ref{3}). Hence, the~integrability of $H_0$ provides a sufficient condition for the validity of the trans-Planckian censorship~conjecture.

In other words, under  assumption (\ref{hub1}), inequality (\ref{1}) (and similarly (\ref{1'})) \emph{implies} (\ref{2}) (or (\ref{3})), but~\emph{not} vice-versa. Sometimes  a stronger version of the trans-Planckian censorship conjecture is stated in the form of a double implication,  which, however, assumes more than just the integrability of $H_0$.  The~following \emph{equivalence},
\be\label{tc0}
\frac{a(t_f)}{a(t_i)}\,l_P<H_f^{-1}\quad \textrm{if and only if}\quad H_0(t_f-t_i)<\ln\frac{M_P}{H_f},
\ee
is true (not just as a one-way implication), provided that the \emph{equality} $\dot{a}/a=H_0$ is assumed  instead of the differential inequality (\ref{hub1}).

Another possible form is to take the  trans-Planckian censorship conjecture to mean the reverse  statement, namely  that $(\ref{2})\Rightarrow(\ref{1})$ for any integrable $H_0$; namely, that for any $t_f$ and any nonzero $H_f$, we have~\cite{tcc1}:
\be\label{tc1}
\frac{a(t_f)}{a(t_i)}\,l_P<H_f^{-1}\quad \textrm{implies}\quad \int_{t_i}^{t_f}H_0(s)ds<\ln\frac{M_P}{H_f}.
\ee
This statement  is different in meaning from  Equations (\ref{2}), (\ref{3}), or~(\ref{tc0}), and~is true provided again that $H_0$ is an integrable~function.

\section{A Breakdown~Criterion}\label{sec3}
In this Section we show that a trans-Planckian bound together with the additional assumption of the existence of a  lower bound for the scale factor are sufficient conditions for producing singularity-free~universes.

First we show that since any of the trans-Planckian bounds discussed in the previous   section provide an upper bound for $a$, we can obtain a criterion about the possible absence of blowup solutions for the scale factor $a$ in any interval of the form $[t_i,t_f]$.

For an initial time $t_i$, we take the the `initial datum' to be $a(t_i)=a_i$, and~consider the maximal interval of existence of solutions $a(t)$ to be $I=(T_-,T_+)$ where $-\infty\leq T_-<t_i< T_+\leq \infty$. Any trans-Planckian bound provides a suitable upper bound for $a$, and~therefore  by the Picard existence and uniqueness theorem (cf. e.g.,~\cite{tao}, p. 14) we have a global solution, which is $T_+=\infty$, that does not go to infinity in a finite time in the~future.

A physical interpretation of this result is that singularities of the finite-time blow-up type for $a(t)$ are strictly prohibited when (\ref{gr1}) holds and $H_0$ is~integrable.

However, in~general relativity a singularity is defined as geodesic incompleteness~\cite{he}.
The previous  discussion does not of course prove geodesic completeness, and~so cannot provide an argument for a resolution of singularities of  spacetime under the above assumptions. The~physical problem is to prove the existence for an infinite proper time, and~in this respect the work in Ref.~\cite{ycb-c} becomes~relevant.

In~\cite{ycb-c}, a~theorem was proven giving sufficient conditions for geodesic completeness in the following sense. We assume the standard $(3+1)$-splitting of a globally hyperbolic spacetime where the lapse function, shift vector field and spatial metric are all bounded (regular hyperbolicity). If~we further take the norms of the spatial gradient of the lapse function as well as that of the extrinsic curvature to be bounded by integrable functions on the interval $[t_i,\infty)$, then it follows that the spacetime is future timelike and null geodesically~complete.

For example, in~the case of an FRW universe with scale factor $a$, the~lapse $N=1$, the~shift $\beta=0$, and~thus the gradient of the lapse vanishes, while the norm of the extrinsic curvature is given by ${|K|_{g}} ^{2}=3({\dot{a}/a})^{2}=3H^{2}$. Hence, this result tells us that in FRW universes that have 
 their scale factor bounded below will be singular only if there is a finite time $t_1\in [t_i,\infty)$ such that the Hubble parameter $H$ is not integrable on the corresponding interval $[t_1,\infty)$.

Previously we assumed the Gronwall bound (\ref{gr1}) for $H$,  where $H_0$ could also be negative, and we discussed its importance in the formulation of trans-Planckian bounds.  That discussion provides only half of the conditions needed for a complete singularity resolution, however, and~we will now discuss the other~half.

Let us introduce the following `anti-Gronwall' assumption, namely,
\be \label{anti}
H(t)\geq b>0,
\ee
with $t\in[t_i,t_f]$, for~some constant $b$, so that
$0<b\leq\dot{a}/a.$
Integrating on $[t_i,t_f]$ we find that,
\be\label{fut-a}
a(t_f)\geq a(t_i)\,e^{b(t_f-t_i)},
\ee
i.e., 
 the scale factor $a$ is bounded from below. This is a way to  circumvent the singularity at $a(t)=0$ for some $t$ earlier than $t_i$ that is expected from the Raychaudhuri equation, because~the anti-Gronwall condition  (\ref{anti}) is the opposite of the usual one, i.e.,~negative expansion (or positive convergence) assumed   in the singularity theorems (cf.~\cite{he}, Thm. 3, p. 271).

The question is then whether the interval $I=(T_-,T_+)$, where the scale factor $a$ is bounded, is finite or infinite. From~the results above it follows that using the anti-Gronwall condition  (\ref{anti}) (i.e., $a$ is bounded below) together with  the trans-Planckian bound,  we find that the \emph{norm} $|H(t)|$ will be bounded  for all  time, not just $H$,  so that the interval $I$ can be infinite (to the left, right, or~both).  This is so because according to the  completeness theorem of~\cite{ycb-c} mentioned above, the~integrability of $|H|$, i.e.,~$|H|$ is bounded by the integrable function $H_0$ as in (\ref{hub1}), and is also a \emph{sufficient} condition for geodesic completeness (the others being that spacetime is globally and regularly hyperbolic) to the past, future, or~both.

We note  that this argument is independent of the the usual assumption on the Ricci tensor, because~the positive convergence condition is an independent hypothesis (i.e., $R^{\mu\nu}X_\mu X_\nu\geq 0$ for non-spacelike vector fields), and~leads to the absence of past or future~singularities.

For a Friedman universe, in~particular,  geodesic completeness can only fail if there  exists a time $t_i$ such that the norm of the Hubble parameter $|H|$ becomes non-integrable in the interval $[t_i,\infty)$. The~non-integrability of $|H|$ provides the only necessary condition for a Friedman universe to be singular. There are different ways for this non-integrability to arise, and~an exhaustive classification of the nature of possible singularities that occur this way was presented in~\cite{co-kl,co-kl2}.

Therefore we are led to  conclude that using the completeness theorem, the~trans-Planckian bounds, and~the anti-Gronwall assumption, there is  a way out of the inevitability of the singular nature of Friedman universes either in the past or future, by~providing conditions for the norm of the Hubble parameter to be bounded and hence be~integrable.

This argument also explains why the trans-Planckian censorship conjecture favors scenarios such as the ekpyrotic  universe where the scale factor is bounded below, or~the emergent universe scenarios~\cite{el-maa} where $H$ is not only integrable but in fact  is asymptotically vanishing~\cite{co-kl},  rather than an inflationary universe where there is a singularity with a \emph{finite} $H$, cf.~\cite{bor,co-kl}.

Therefore, we conclude that future (or past) geodesic completeness and the associated absence of future (past) singularities is a necessary consequence of trans-Planckian bounds in any scenario in which the universe satisfies the anti-Gronwall~assumption.

\section{Examples}
As an application of the previous results, we consider here a few representative examples that illustrate some of the features of the use of trans-Planckian bounds in proving geodesic~completeness.

A a first example, let us consider the emergent universe scenario of~\cite{el-maa}.  For~this model, the~Gronwall hypothesis, namely that the expansion is sub-Hubblian, together with the trans-Planckian bound (\ref{2}), implies that  the initial (Einstein static universe) scale factor $a(t_i)$ is bounded from below, avoiding the usual fine-tuning issues associated with the emergent scenario. In~addition, the~anti-Gronwall bound on the Hubble parameter (\ref{anti}) implies a large classical expanding universe with a scale factor given by (\ref{fut-a})  at late times. This universe is also future geodesically complete because the Hubble parameter is not only bounded by asymptotically vanishing, cf.~\cite{co-kl}.

In fact it is not difficult to devise universes with an asymptotically vanishing Hubble parameter, thus signaling future geodesic completeness. As~an example, in~any flat or negatively curved FRW model filled with a perfect fluid and  scalar field  with a positive, bounded potential, one can show that not only $H$ but also the fluid density are future asymptotically vanishing, cf.~\cite{mir6}, Proposition 2. Hence, in~any model with  logarithmic or generalized potentials, e.g.,~of the form studied in~\cite{bap1,bap2}, the~trans-Planckian bound together with the anti-Gronwall hypothesis imply a singularity-free~evolution.

\section{Discussion}
In this paper we have discussed the role of trans-Planckian bounds in relation to the formation of singularities. We have first shown that such bounds can be naturally deduced from the Gronwall hypothesis, which  provides upper bounds to the Hubble~parameter.

This leads to a new criterion for the  absence of diverging cosmological solutions either at a finite time or at~infinity.

Furthermore, we have shown that trans-Planckian bounds, when combined with the condition that the Hubble parameter is bounded away from zero, lead to geodesically complete universes satisfying  the usual causality assumptions.
We therefore conclude that trans-Planckian bounds provide a way to singularity-free universes if the Hubble parameter is~integrable.

This result opens the way to constructing singularity-free cosmologies starting from a trans-Planckian bound and examining the integrability of the expansion parameter. This in turn depends on the type of matter content of the universe, and~may lead to selection rules for non-singular cosmologies from suitable restrictions on the fluid or other parameters of the matter fields. Due to the generality of our criteria, we believe that our present results may also  be extended to more general  anisotropic or inhomogeneous~cosmologies.

\section*{Acknowledgments}S.C. is grateful to Robert Brandenberger for valuable comments on an earlier version of this work. We thank David Andriot and Yong Cai for useful~correspondence. The research of S.C. was funded by scientific project FSSF-2023-0003.


\end{document}